\documentclass[reprint,
superscriptaddress,
altaffillsymbol,
groupedaddress,
amsmath,amssymb,
aps,
prb,
]{revtex4-2}
\usepackage{graphicx}% Include figure files
\usepackage{dcolumn}% Align table columns on decimal point
\usepackage{bm}% bold math
\usepackage{dsfont}
\usepackage{physics}
\usepackage{color}
\usepackage{soul}
\setstcolor{red}

\usepackage{ifthen}
\newboolean{showannotations}   
\setboolean{showannotations}{false} 
\newcommand*{\added}[1]{%
  \ifthenelse{\boolean{showannotations}}{{\color{blue}#1}}{#1}%
}

 \usepackage[thinc]{esdiff}

\newcommand*{\removed}[1]{%
  \ifthenelse{\boolean{showannotations}}{\st{#1}}{}%
}

\newcommand*{\change}[2]{%
  \ifthenelse{\boolean{showannotations}}{{\color{red}\st{#1}}{\color{blue}#2}}{#2}%
}

\begin{document} 
% Force * for email and numbers for others

\title{  Microscopic Theory of Polaron-Polariton Dispersion and Propagation}
\author{Logan Blackham}
\altaffiliation{equal contribution}
\affiliation{Department of Chemistry, Texas A\&M University, College Station, Texas 77843, USA}
\author{Arshath Manjalingal}%
\altaffiliation{equal contribution}
\affiliation{Department of Chemistry, Texas A\&M University, College Station, Texas 77843, USA}

\author{Saeed Rahmanian Koshkaki}%
\affiliation{Department of Chemistry, Texas A\&M University, College Station, Texas 77843, USA}
\author{Arkajit Mandal}%
\email{mandal@tamu.edu}
\affiliation{Department of Chemistry, Texas A\&M University, College Station, Texas 77843, USA}

\begin{abstract}
We develop an analytical microscopic theory to describe the polaron-polariton dispersion, formed by hybridizing excitons, photons, and phonons, and their coherent dynamics inside optical cavities. Starting from a microscopic light-matter Hamiltonian, we derive a simple analytical model by pursuing a non-perturbative treatment of the phonon and photon couplings to excitons. Within our theoretical framework, the phonons are treated as classical fields that are then quantized via the Floquet formalism. We show that, to a good approximation, the entire polaron-polariton system can be described using a band picture despite the phonons breaking translational symmetry. Our theory also sheds light on the long-lived coherent ballistic motion of exciton-polaritons with high excitonic character that propagate with group velocities lower than is expected from pure exciton-polariton bands, offering a microscopic explanation for these puzzling experimental observations.  
\end{abstract}

\maketitle
{\footnotesize

\textbf{Introduction.} Coupling quantized electromagnetic radiation to excitons forms exciton-polaritons (EPs), a hybrid photon-matter quasi-particle~\cite{LiARPC2022, SnokePhyToday2010, DengRMP2010, LidzeyPRL1999, SandikNatMat2024, KowalewskiJCP2016, MandalCR2022}, that demonstrates a wide range of exotic phenomena~\cite{KeelingARPC2020, YueningACSP2024,KockumNRP2019, XuNN2025, Wurdack2023NC, Matthias2021NC, SuyabatmazJCP2023, LiuNP2015, kenaNP2010}, including enhanced transport surpassing the inherent limits of bare-exciton transport~\cite{StegerPRB2013, XuNC2023, BalasubrahmaniyamNM2023, SandikNatMat2024, sanvittoNM2016, FerreiraPRM2022, DhamijaCPR2024}. This extraordinary phenomenon, namely cavity-enhanced exciton transport, demonstrates the unique nature of exciton-polaritons, redefining the traditional paradigms of energy transport with possible applications in quantum information science and chemical reactivity~\cite{SandikNatMat2024, MandalCR2022, KadeCR2020}. 

A superposition of neighboring exciton states in reciprocal space leads to coherent ballistic propagation with a group velocity equal to the slope of the band structure in the absence of dissipation~\cite{Reineker1984Book}. Phonons, which are intrinsic to materials, break the translational symmetry of an excitonic system, leading to phonon-induced decoherence and incoherent diffusive motion~\cite{TroisiPRL2006, Jonathan2020PRX, Linjun2011JCP, YarkonyJCP1977}. Therefore, it is expected that the coherent ballistic motion of exciton-polaritons will exhibit group velocities matching the exciton-polariton dispersion for times less than the decoherence lifetime~\cite{BalasubrahmaniyamPRB2021, MandalCR2022, Krupp2024}. Interestingly, recent experiments~\cite{XuNC2023, BerghuisACSp2022, PandyaAdvS2022, BalasubrahmaniyamNM2023} indicate that exciton-polaritons with significantly high excitonic character (up to $\sim$50\% excitonic) show long-lived coherent ballistic motion (up to hundreds of femtoseconds)~\cite{StegerPRB2013, LiuACSP2020, XuNC2023, BalasubrahmaniyamNM2023} with group velocities lower than the slopes of the exciton-polariton band structure~\cite{XuNC2023, BalasubrahmaniyamNM2023,PandyaAdvS2022, SandikNatMat2024}. Despite many recent insightful theoretical works on exciton-polariton dynamics~\cite{XuNC2023, SokolovskiiNatCom2023, TichauerAS2023, ChngNL2025, Ying2024, TutunnikovArxiv2024, AroeiraNp2024, EngelhardtPRL2023}, including a recent inspiring work~\cite{Ying2024} focusing on the group velocity renormalization phenomena within a perturbative framework, a full microscopic understanding of this extraordinary phenomenon has remained elusive. %In particular, a recent inspiring work~\cite{Ying2024} employed a perturbative approach to understand this group velocity renormalization. 

% which breaks down at high wavevector $k$ where it predicts negative group velocities. 
%The ambiguity of phonon-exciton-polariton interactions has proven difficult to justify and quantify, and specifically, it has remained a mystery as to why the group velocities of exciton-polaritons are {\it renormalized} in the presence of phonons~\cite{XuNatCom2023, SandikNatMat2024}.

 Here we introduce a new theoretical framework to understand the complex polariton dispersion formed by hybridizing excitons, photons, and phonons, as well as their coherent dynamics inside optical cavities. Given the intractable nature of the full quantum mechanical problem, we introduce a convenient picture where exciton-polaritons are embedded in a classical phonon field. We quantize this phonon field using the Floquet formalism to derive an analytical model exhibiting {\it translational} symmetry to a good approximation, allowing for coherent motion. This analytical model produces an extremely accurate description of exciton-polariton dispersion when compared to the angle-resolved polariton spectra obtained using a mixed quantum-classical approach~\cite{TichauerAS2023, XuNC2023, Ying2024, ChngNL2025,SokolovskiiNatCom2023}. Using our model, we show that the presence of phonons introduces vibronic structure in the exciton-polariton dispersion, which we refer to as the polaron-polariton dispersion. We show that this vibronic structure is responsible for a renormalization of the group velocity and that despite a strong interaction with phonons, an effective band structure model can be adopted. Our theory not only serves as a convenient analytical model to understand polariton spectra but also provides new insights into the interplay between phonons and exciton-polaritons. 

% I like this part, maybe we can work it into the end if we need a new ending [[We show that a massive simplification can be made allowing for the extraction of the quantum vibronic structure for a system derived with classical limits, outlining the true effect phonons have on cavity-enhanced exciton transport within a basis a fraction of the size.]]

{\bf Theory}. We consider a  generalized multimode Holstein-Tavis-Cummings Hamiltonian ~\cite{KeelingARPC2020, MandalCR2022, MandalNL2023, SokolovskiiNatCom2023}, which describes an exciton-polariton system beyond the long-wavelength approximation, interacting with phonons and is written as

\begin{align}
\hat{H}_{\mathrm{LM}} &=  \sum_n \hat{X}_{n}^{\dagger} \hat{X}_{n} \varepsilon_{0} + \sum_k \hat{a}_{k}^{\dagger}\hat{a}_{k} \omega_{c}(k) + \sum_{n} \frac{\hat{P}_{n}^2}{2} + \frac{1}{2}\omega^2 \hat{R}^{2}_n \nonumber \\
&+ \tau \sum_n(\hat{X}_{n}^{\dagger} \hat{X}_{n+1} + \hat{X}_{n+1}^{\dagger} \hat{X}_{n}) + \sum_n \gamma \hat{X}_{n}^{\dagger} \hat{X}_{n} \hat{R}_{n} \nonumber \\
&+   \sum_{n,k} \frac{ \Omega_k}{\sqrt{N}}\bigg[\hat{a}^{\dagger}_{k}\hat{X}_{n}e^{-i k  \cdot r_{n}} + \hat{a}_{k}\hat{X}_{n}^{\dagger}e^{i k \cdot r_{n}} \bigg].  \end{align}
Here $\hat X^\dagger_{n}$ ($\hat a^\dagger_{k}$) creates an excitation (photon) at site $n$ (mode $k$),  and ${R_n}$ (${P_n}$) is the position (momentum) operator for the $n$th phonon mode. Here $\varepsilon_{0}$ is the on-site energy with each site located at $r_n = a\cdot n$ with $a$ as the lattice constant, $\tau$ is the hopping parameter, $\gamma$ is the exciton-phonon coupling, and $\Omega_k = \Omega \sqrt{\omega_0/\omega_{c}(k)}$ is the exciton-photon coupling. Finally, $\omega_{c}(k)$ and $\omega$ are the photon and phonon frequency, respectively. Further details are provided in the supporting information. Notably, the phonon degrees of freedom break the translational symmetry of the exciton-polariton system. Consequently, the polaron-polariton, formed through the hybridization of excitons, photons, and phonons, is not expected to exhibit a strict band structure. Nevertheless, we demonstrate that a quasi-band structure framework can be employed, effectively capturing the complex ballistic transport of exciton-polaritons.

Direct (analytical or numerical) quantum mechanical treatment of this light-matter Hamiltonian is a formidable task given that polaritonic dispersion can only be obtained when using $N \sim 10^{5}$ sites for experimentally relevant values of the lattice constant $a$ (chosen here to be $1.2~nm$). To solve this intractable problem, we employ a mixed-quantum-classical approach, namely the mean-field Ehrenfest (MFE) method~\cite{MandalCR2022, Hoffmann2020JCP, LiPRA2018}, that is known to accurately reproduce quantum vibronic structure in optical spectra in a single-site exciton-phonon model~\cite{EgorovJPCB1999, PetitJCP2014}, despite the classical treatment of phonons. Within this approach, the phonon modes are treated classically, i.e. $\{\hat{R}_n, \hat{P}_n\} \rightarrow \{{R}_n, {P}_n\} $, while the photonic and excitonic parts are propagated quantum mechanically using the {\it polaritonic} Hamiltonian $\hat{H}_{\mathrm{pl}} ({\bf R}) =  \hat{H}_{\mathrm{LM}} - \sum_{n} {{P}_{n}^2/2} - \omega^{2} R_n^2/2 $. The equations of motion in the MFE approach (in atomic units) are  written as
\begin{align}
i|\dot{\Psi} (t)\rangle &= \hat{H}_{\mathrm{pl}} ({\bf R})|{\Psi (t)}\rangle, \label{TDSE} \\
\ddot{R}_n(t) &= \dot{P}_n(t) =  - \Big\langle {\Psi} (t)\Big| \diff{\hat{H}_{\mathrm{LM}} ({\bf R})}{R_n} \Big |{\Psi}(t)\Big\rangle.
\end{align}
The initial nuclear coordinates $\{{R}_n(0), {P}_n(0)\}$ are sampled from a Wigner distribution (see details in the Supporting Information), and an expectation value of an operator $\hat{A}$ is computed as $\langle \hat{A} \rangle \approx \big\langle \langle {\Psi} (t)|\hat{A}| {\Psi} (t)\rangle \big\rangle_{\mathrm{MFE}}$ where $\langle ... \rangle_{\mathrm{MFE}}$
indicates averaging over realizations of initial nuclear coordinates $\{{R}_n(0), {P}_n(0)\}$.

\begin{figure}
\centering
\includegraphics[width=1.0\linewidth]{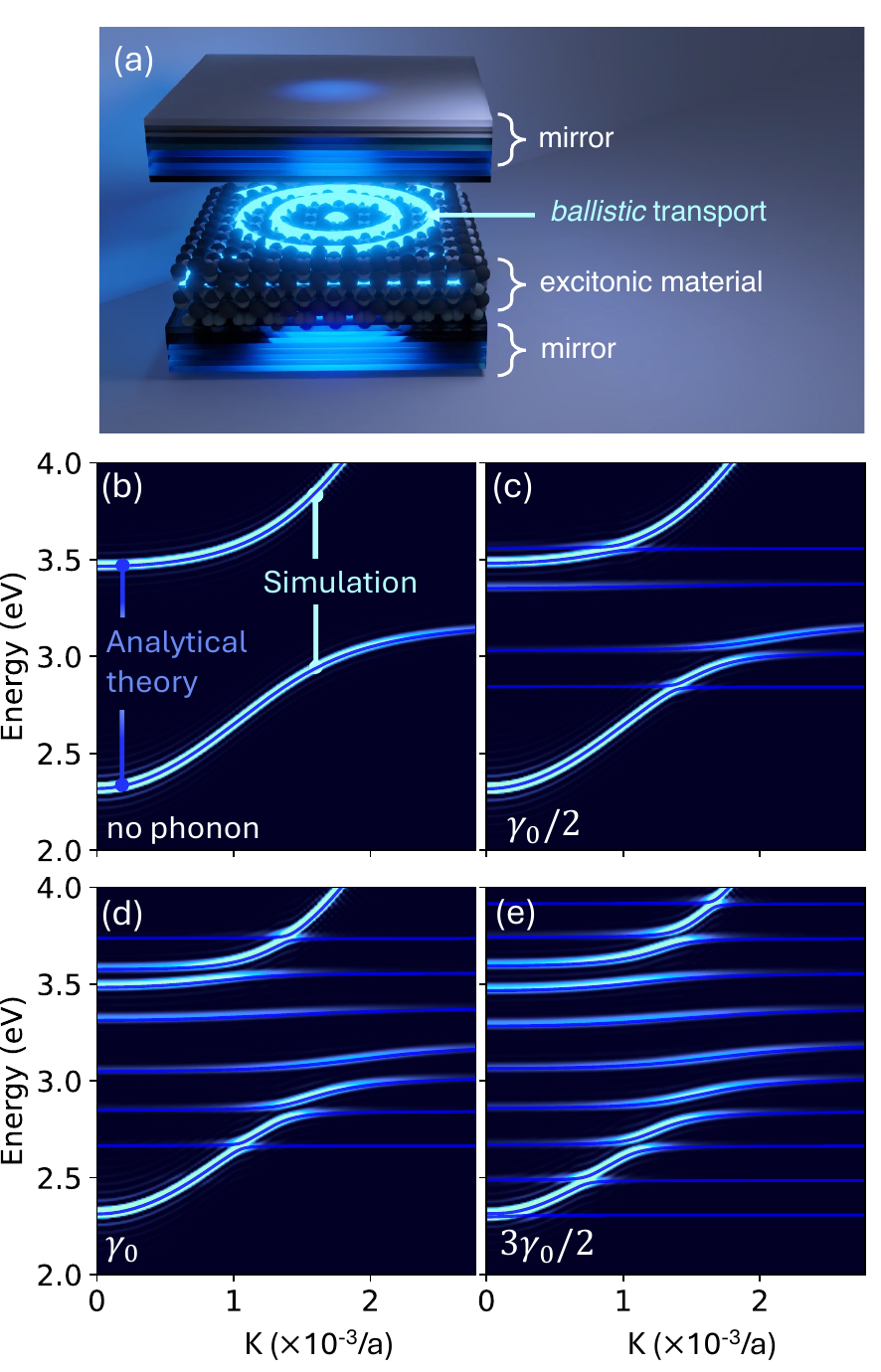}
\caption{\footnotesize (a) Schematic 3-D model of exciton-polariton transport within an optical cavity. (b) Exciton-polariton band structure from simulation and theory with no phonon coupling, (c) with phonon coupling $\gamma_{0}/2$,  (d) with phonon coupling $\gamma_{0}$, (e) with phonon coupling $3\gamma_{0}/2$, where $\gamma_{0}$ is the phonon coupling. The parameter $\gamma_{0} = 5.85 \times 10^{-4}$ a.u. Further we use $\Omega = 3900$  cm$^{-1}$, $N = 40001$, $\tau = 0$, $\omega_0 = 2.58$ eV, and $\varepsilon_0 = 3.2$ eV.}
\label{fig1}
\end{figure}

The angle-resolved optical spectra $I(\omega, k)$ can be obtained by directly propagating the quantum dynamics to compute  
\begin{align}
I(\omega, k) = \lim_{T\rightarrow \infty} \int_{0}^T dt~ e^{i \omega t} \big\langle \langle 1_k| {\Psi} (t)\rangle \big\rangle_{\mathrm{MFE}} \cdot \cos({\pi t}/{2T}),
\end{align}
where $| {\Psi} (0)\rangle = \hat{a}_k^{\dagger} |\bar{0}\rangle = |1_k\rangle$. Note that we have added the term $\cos ({\pi t}/{2T})$ to suppress spurious Gibbs oscillations. Our numerical result, presented in Fig.\ref{fig1}, illustrates the emergence of complex vibronic structure in the momentum-resolved polaritonic spectra in the presence of phonon modes. As can be seen in these figures, despite the absence of a strict translational symmetry, the angle-resolved spectra suggest the existence of a quasi-dispersion of polaron-polaritons. Such vibronic structure in exciton-polariton bands has been seen in recent experiments~\cite{DanielCS2020, DonghaiPRL2022}. Below, we derive the analytical forms of these quasi-bands with details provided in the Supporting Information. 

To obtain an analytical expression for these polaron-polariton (quasi) bands, we first make the classical path approximation~\cite{AkimovJCTC2013,Linjun2011JCP,YamijalaJPCA2021}, such that $\ddot{R}_n(t) \approx  - \omega^2 {R}_n(t)$ with  
\begin{align}
{R}_n(t) &\approx {R}_n(0) \cos\omega t + \frac{1}{\omega}{P}_n(0) \sin\omega t .
\end{align}
With this analytical expression of ${R}_n(t)$, the dynamics of the exciton-polariton wavefunction $|\Psi\rangle$ can be thought to be evolving under the time-periodic Hamiltonian $\hat{H}_{\mathrm{pl}} (t)$ expressed as
\begin{align}
\hat{H}_{\mathrm{pl}}(t) = \hat{H}_{\mathrm{EP}} + \hat{P} e^{i \omega t} + \hat{P}^{\dagger} e^{-i \omega t} ,
\end{align}
where $ \hat{P}  = \sum_n \gamma \hat{X}_{n}^{\dagger} \hat{X}_{n} Z_{n} 
$ describes the interaction to a phonon field with $Z_{n} = {R_{n}(0)}/{2} + {P_{n}(0)}/{2i\omega}$ and $\hat{H}_{\mathrm{EP}}$ as the pure exciton-polariton Hamiltonian written as

\begin{align}
\hat{H}_{\mathrm{EP}} &=  \sum_n \hat{X}_{n}^{\dagger} \hat{X}_{n} \varepsilon_{0}  + \tau \sum_n(\hat{X}_{n}^{\dagger} \hat{X}_{n+1} + \hat{X}_{n+1}^{\dagger} \hat{X}_{n}) \nonumber \\
&+\sum_k \hat{a}_{k}^{\dagger}\hat{a}_{k} \omega_{c}(k) +   \sum_{n,k} \frac{ \Omega_k}{\sqrt{N}} \bigg[\hat{a}^{\dagger}_{k}\hat{X}_{n}e^{-i k  \cdot r_{n}} + \hat{a}_{k}\hat{X}_{n}^{\dagger}e^{i k \cdot r_{n}} \bigg] \nonumber.
\end{align}
Notice the similarity between $\hat{H}_{\mathrm{pl}} (t)$ and the typical laser-matter Hamiltonian, with phonon degrees of freedom (or molecular vibrations) in our system playing the same role as a laser field.  We adapt the Floquet formalism~\cite{Shirley1965PR, TaylorCPR2024} and rewrite  $\hat{H}_{\mathrm{pl}}(t)$  in an extended space (so-called Sambe space) as a time-independent Hamiltonian $\hat{H}_{\mathrm{F}}$ such that
\begin{align}
\hat{H}_{\mathrm{pl}}(t) \mapsto \Hat{{H}}_{\mathrm{F}} &= \lim_{M\to\infty} \sum_{ij} \mathbb{P}_{j} \hat{\mathcal{H}}_{\mathrm{F}} \mathbb{P}_{i}, \nonumber\\
\textrm{with}~~  \mathbb{P}_{i} &\in    \Big\{  \hat{X}_{n}^{\dagger} \frac{(\hat{B}^{\dagger})^{M+m}}{\sqrt{(M+m)!}}|\bar{0} \rangle, \hat{a}_{k}^{\dagger} \frac{(\hat{B}^{\dagger})^{M+m}}{\sqrt{(M+m)!}}|\bar{0} \rangle \Big\}. 
\end{align}
Here we have introduced the bosonic operator $\hat{B}$ that creates an excitation in the phononic field. Further, $\hat{\mathcal{H}}_{\mathrm{F}}$ is expressed as
\begin{align}
\hat{\mathcal{H}}_{\mathrm{F}} &= \sum_{n} \Big(\varepsilon_{0}  + \dfrac{\gamma}{\sqrt{M}} ({Z}_{n}\hat{B} + {Z}_{n}^{*}\hat{B}^{\dagger}) \Big)\hat{X}_{n}^{\dagger}\hat{X}_{n}  \nonumber \\ 
 &+ \tau \sum_n(\hat{X}_{n}^{\dagger} \hat{X}_{n+1} + \hat{X}_{n+1}^{\dagger} \hat{X}_{n}) +  (\hat{B}^{\dagger}\hat{B}- M)\omega + \sum_{k} \hat{a}_{k}^{\dagger}\hat{a}_{k}\omega_{c}(k)\nonumber\\
& +  \sum_{n,k} \frac{ \Omega_k}{\sqrt{N}} (\hat{a}^{\dagger}_{k}\hat{X}_{n}e^{-i k  \cdot r_{n}} + \hat{a}_{k}\hat{X}_{n}^{\dagger}e^{i k \cdot r_{n}}).
\end{align}

Next, we perform a polaron transformation on $\hat{\mathcal{H}}_{\mathrm{F}}$ using the operator $\hat{U}_{D}$ defined as
\begin{align}
\hat{{U}}_{D} =  \prod_n{  \exp \Bigg[\Big({ Z_{n}^*\hat{B}^{\dagger} - Z_{n}\hat{B}} \Big) \dfrac{\gamma \hat{X}_{n}^{\dagger}\hat{X}_{n}}{\omega\sqrt{M}}\Bigg]}
\end{align}
to obtain $\hat{\mathcal{H}}_{\mathrm{F}}'  = \hat{U}_{D}^{\dagger} \hat{\mathcal{H}}_{\mathrm{F}} \hat{U}_{D}$ that is explicitly written as

\begin{align}
&\hat{\mathcal{H}}_{\mathrm{F}}' = \sum_{n} \varepsilon_{0} \hat{X}_{n}^{\dagger}\hat{X}_{n} +  (\hat{B}^{\dagger}\hat{B}- M)\omega + \sum_{k} \hat{a}_{k}^{\dagger}\hat{a}_{k}\omega_{c}(k)  \\
&+ \tau \sum_n\Bigg(\hat{X}_{n}^{\dagger} \hat{X}_{n+1}  \exp\Big[{\gamma\dfrac{\Delta Z_{n} \hat{B} - \Delta Z_{n}^{*}  \hat{B}^{\dagger}}{\omega\sqrt{M}}}\Big] + h.c.\Bigg) \nonumber\\
&+  \sum_{n,k}  \frac{ \Omega_k}{\sqrt{N}} \Big(\hat{a}^{\dagger}_{k}\hat{X}_{n}\exp\Big[{\dfrac{\gamma(Z_{n}\hat{B} - Z_{n}^{*}\hat{B}^{\dagger})}{\omega\sqrt{M}}  } -i k \cdot r_{n} \Big] + h.c.\Big), \nonumber
\end{align}
where $\Delta Z_{n} = Z_{n+1} - Z_{n}$. To arrive at a simpler form we further restrict our subspace such that  $ \mathbb{P}_{i} \in \mathcal{S} =   \big\{  \hat{X}_{n}^{\dagger} \frac{(\hat{B}^{\dagger})^{M+m}}{\sqrt{(M+m)!}}|\bar{0} \rangle, \hat{a}_{k}^{\dagger} \frac{(\hat{B}^{\dagger})^{M}}{\sqrt{M!}}|\bar{0} \rangle \big\}  $ with $M\rightarrow \infty$.  That is, here we only consider the reference excitation block, the $M$th block, of the phonon field for states with a single photon. Further, we adapt the simplified notation $|n, m\rangle \equiv \lim_{M\to\infty} 
 \hat{X}_{n}^{\dagger} \frac{(B^{\dagger})^{M+m}}{\sqrt{(M+m)!}}|\bar{0} \rangle$ and $ |1_k\rangle \equiv \lim_{M\to\infty} \hat{a}_{k}^{\dagger} \frac{(B^{\dagger})^{M}}{\sqrt{M!}}|\bar{0} \rangle$ and obtain

\begin{align}
\Hat{{H}}_{\mathrm{F}}' &= \lim_{M\to\infty} \sum_{ij\in \mathcal{S}} \mathbb{P}_{j} \hat{\mathcal{H}}_{\mathrm{F}}' \mathbb{P}_{i} = \lim_{M\to\infty} \sum_{ij\in \mathcal{S}} \mathbb{P}_{j} (\hat{U}_{D}^{\dagger} \hat{\mathcal{H}}_{\mathrm{F}} \hat{U}_{D}) \mathbb{P}_{i} \label{real-space-h}\\
&= \Bigg[ \sum_{k} \omega_{c}(k) \big| 1_k\big\rangle \big\langle  1_k\big| +  \sum_{n, m} \big(\varepsilon_{0} + m\omega\big)~\big|n, m\big\rangle \big\langle n, m\big| \nonumber \\
&+  \tau \sum_{m, m',n} Q_{mm'}(\Delta Z_n)\Big(\big|n, m \big\rangle \big\langle n+1, m'\big| + h.c.\Big) \nonumber\\
&+   \sum_{m,n,k} \frac{ \Omega_k}{\sqrt{N}}\Big( Q_{m0}(Z_n)e^{-i k \cdot r_{n}}  \big|n, m \big\rangle \big\langle 1_k \big| + h.c.\Big)\Bigg]. \nonumber 
\end{align}
Here $Q_{mm'}(\alpha)=  \lim_{M \rightarrow \infty} \langle m  + M | e^{\frac{\gamma(\alpha\hat{B} - \alpha^{*}\hat{B}^{\dagger})}{\omega\sqrt{M}}  }   |m' + M\rangle $ is the overlap between infinitely excited displaced harmonic oscillator states. In order to obtain the polariton quasi-band structure, we introduce the following {\it effective} reciprocal exciton states,
\begin{align}\label{effective}
\big|k, m \big\rangle =  \sum_{n} \frac{Q_{m0}(Z_n)} {\sqrt{\mathcal{S}_m}} e^{-i k \cdot r_{n}}  \big|n, m\big\rangle \equiv \hat{Y}^{\dagger}_{k,m} |\bar{0}\rangle ,
\end{align}
where $\mathcal{S}_m = \sum_n Q_{m0}^2(Z_n)$ is a normalization factor. 
 Next, we approximate $\{\big|k, m \big\rangle \}$ to be orthogonal to each other (see details in the SI), which holds true for $k \rightarrow 0$ where the dynamics of the exciton-polariton is confined due to the sharp band structure of the photon. Further, we replace the third line of Eq.~\ref{real-space-h} with the expectation value $ \big\langle k, m \big| \tau \sum_{m', m'',n} Q_{m'm''}(\Delta Z_n)\Big(\big|m', n\big\rangle \big\langle m'', n+1\big| + h.c\Big) \big|k, m \big\rangle \approx  2 \tau \sum_n Q_{00}(\Delta Z_{n}) \cdot  Q_{m0}(Z_n) \cdot Q_{m0}(Z_{n+1})/\mathcal{S}_m = 2 \tau \cdot \xi_0 $ to obtain the final form of our model as

\begin{align}\label{HFeff}
 \Hat{{H}}_{\mathrm{F}}' \approx  \sum_{k}& \Bigg[ \hat{a}_{k}^{\dagger} \hat{a}_k\omega_{c}(k)  + \sum_{m} ({\varepsilon}_{0} + 2\tau \xi_0 + m\omega)\hat{Y}^{\dagger}_{k,m} \hat{Y}_{k,m} \nonumber \\
+& \sum_{m}   \sqrt{\frac{\mathcal{S}_m}{N}} \Omega_k   \Big(\hat{Y}^{\dagger}_{k,m} \hat{a}_{k} + \hat{a}_{k}^{\dagger} \hat{Y}_{k,m}\Big)\Bigg] =  \sum_k \Hat{\mathcal{H}}_k
\end{align}
which is block diagonal in each $k$, thus allowing us to extract the phonon-modified exciton-polariton (or equivalently polaron-polariton) dispersion.  The polaron-polariton (quasi) bands are obtained by diagonalizing $\Hat{\mathcal{H}}_k$ written as

\begin{align}\label{model}
 \Hat{\mathcal{H}}_k =  
 \begin{bmatrix}    
 \ddots & \vdots & \vdots & \vdots &   & \vdots\\
 \hdots & \bar{\varepsilon} -\omega  & 0 & 0 &  \hdots & \sqrt{{\frac{\mathcal{S}_{-1}}{N}}}\Omega_k  \\
\hdots & 0 & \bar{\varepsilon}   & 0 &  \hdots & \sqrt{{\frac{\mathcal{S}_0}{N}}}\Omega_k  \\
\hdots & 0 & 0  &  \bar{\varepsilon} + \omega  & \hdots & \sqrt{{\frac{\mathcal{S}_1}{N}}}\Omega_k \\
    & \vdots & \vdots & \vdots & \ddots & \vdots\\
\hdots &   \sqrt{{\frac{\mathcal{S}_{-1}}{N}}}\Omega_k & \sqrt{{\frac{\mathcal{S}_0}{N}}}\Omega_k & \sqrt{{\frac{\mathcal{S}_1}{N}}}\Omega_k & \hdots & \omega_{c}(k) & 
 \end{bmatrix},
\end{align}

where $\bar{\varepsilon} = {\varepsilon}_{0} + 2\tau \cdot \xi_0$. Note that at $\gamma \rightarrow 0$ we have $\mathcal{S}_m  \rightarrow  N\delta_{0m}$ thereby reducing Eq.~\ref{model} to the exciton-polariton band model in the absence of phonons. Overall, we find that  phonon interactions modify the exciton polariton bands in specifically two ways. One, it introduces vibronic states, which are effectively captured via the collective phonon field excitations $\hat{B}^\dagger$ within our mixed quantum-classical framework, coupling to the photonic bands forming a Rabi-Splitting of $\sqrt{\mathcal{S}_m/N}\Omega_k = \sqrt{\sum_n Q_{m0}^2(Z_n)/N}\Omega_k$.  Which can be calculated by integrating the squared overlap $Q_{m0}^2(Z_n)$ over the Wigner distribution of $\{{R}_n(0), {P}_n(0)\}$ such that  
\begin{align}\label{eqnS}
{\mathcal{S}_m} &= \lim_{M\rightarrow\infty} 2\tanh(\frac{\beta\omega}{2})\int_{-\infty}^{\infty}  dz  {\frac{M!}{(M+m)!}}   \Bigg(\frac{\gamma^2 z^2}{\omega^2 M} \Bigg)^{m}     \\
\times&\exp\Big[- \Big(4 \omega \tanh(\frac{\beta\omega}{2}) + \frac{z^2 \gamma^4}{4\omega^4 M^2}\Big) z^2\Big]  \times \Big[ L_{M}^{(m)}(\gamma^2 z^2/\omega^2 M) \Big]^2. \nonumber 
\end{align}

Here $L_{M}^{(m)}$ is the associated Laguerre polynomial. Note that our model presented in Eqn.~\ref{HFeff}-\ref{eqnS}, derived directly from the multi-mode dipole-gauge Hamiltonian, is structurally different from previously proposed models for polaritonic spectra~\cite{spano2015optical, MazzaPRB2013, FontanesiPRB2009} that incorporated the vibronic progression in the polariton dispersion in an ad-hoc manner and effectively assumed translational symmetry a priori. 

%We find that these previous models can not reproduce the spectra or dynamical results presented here. 

%Specifically, Ref.~\ref{MazzaPRB2013, FontanesiPRB2009} assumes that the phonon-exciton-polariton Hamiltonian is 

Second, the phonons renormalize the hopping term $\tau$, simply shifting up the excitonic energy near $k\rightarrow 0$ as expected. In the following, we will focus on the vibronic structure and its implication for the polariton dispersion and set $\tau = 0$ (relevant for molecular exciton-polaritons). In the Supporting Information, we present results when $\tau \neq 0$ relevant for  exciton-polaritons in extended materials. 

{\bf Results and Discussion.} Fig.~\ref{fig1} presents the angle-resolved polariton spectra comparing with polariton (quasi) bands obtained using our analytical model presented in Eq.~\ref{model}. Fig.~\ref{fig1}a schematically illustrates an excitonic material placed inside a Fabry-P\'{e}rot cavity exhibiting ballistic transport of polaritons. The angle-resolved polariton spectra in the absence of phonons, presented in Fig.~\ref{fig1}b reduce to the two-band coupled oscillator model used widely~\cite{KeelingARPC2020, MandalCR2022,DengRMP2010}. Our theoretical model (solid blue lines) reduces to this two-band model when setting $\gamma = 0$ (no phonon interactions), $\mathcal{S}_m = N\delta_{0m}$ thus exactly reproducing the polaritonic spectra in Fig.~\ref{fig1}b. 

Fig.~\ref{fig1}c-e presents the polariton spectra obtained from our direct quantum dynamical simulation at various phonon couplings $\gamma$, comparing them to the predictions of our analytical model. Overall, our analytical model nearly exactly reproduces the polaritonic dispersion obtained numerically,  validating the approximations made in this work to arrive at Eq.~\ref{model}. At relatively small phonon coupling $\gamma = \gamma_0/2$, we observe a clear phonon-induced splitting of the lower polariton band around $3$ eV. Note that with increasing phonon coupling, the exciton on-site energy minima along the phonon displacement coordinate is shifted down by the reorganization energy $\lambda = \frac{1}{2}\frac{\gamma^2}{\omega^2}$. As a result, increasing $\gamma$ also leads to the formation of phonon-induced splitting at progressively lower energy. At the same time,  increasing $\gamma$ also introduces more splitting. This is because the increase in the  displacement of the phonon field leads to  more sizable overlap between the photonic states and excitonic states with $m$ phonon field excitation or de-excitation in the Floquet picture employed here. It is worth noting that while these splittings may look similar to the spectra of a single displaced harmonic oscillator, such a simple model cannot be employed to describe the complex polariton dispersion. The splitting $\propto \sqrt{\mathcal{S}_m} = \sqrt{\sum_n Q_{m0}^2({Z_n})}$ is obtained accurately when including the fluctuations $Z_n$ of all phonon modes in the system and cannot be estimated accurately using some expectation value of $Z_n$ since  $\sqrt{\mathcal{S}_m} \neq \sqrt{Q_{m0}^2(\langle{Z_n}\rangle)}$, where $\langle ... \rangle$ indicates phase space averaging.

\begin{figure}[t]
\centering
\includegraphics[width=1.0\linewidth]{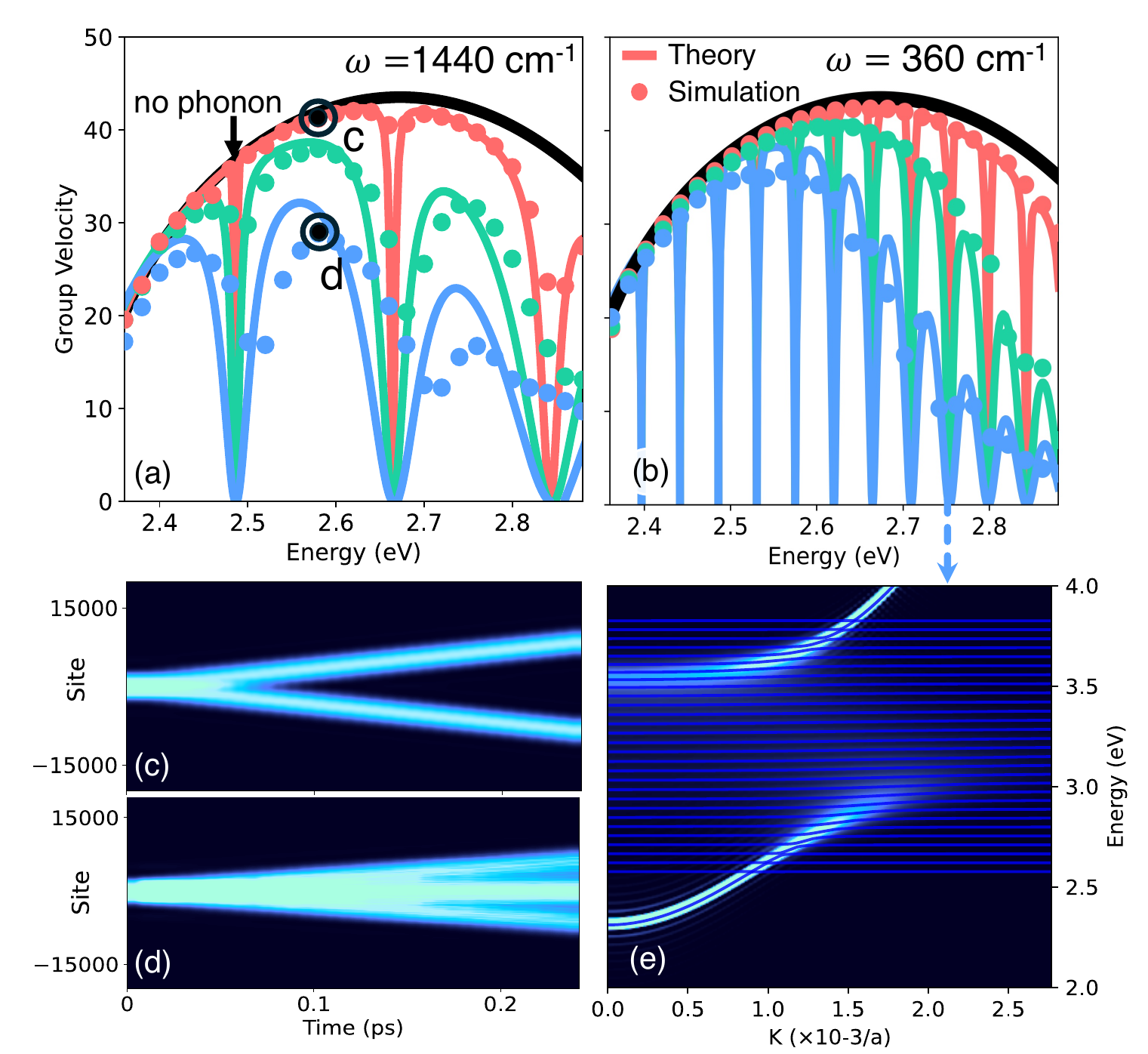}
\caption{\footnotesize  Group velocities extracted from quantum dynamical simulations (filled circles) compared to the predictions of the analytical model (solid lines) introduced in this work, with phonon frequency of (a) $1440 \ \text{cm}^{-1}$ and (b) $360 \ \text{cm}^{-1}$. In (a), the group velocities for different phonon coupling are plotted, with $\gamma_0/2$ represented in red, $\gamma_0$ in green, and $3\gamma_0/2$ in blue. Similarly, in (b), the phonon coupling strengths are depicted as $\gamma_0$ (red), $3\gamma_0/2$ (green), and $2\gamma_0$ (blue).(c),(d) Heatmaps gathered from MFE exciton-polarization propagation over 0.242 ps, corresponding to highlighted points in (Fig 2a). Panels (c) and (d) illustrate the phonon coupling strengths ranging from $\gamma_0/2$ to $3\gamma_0/2$ for a phonon frequency of $1440 \ \text{cm}^{-1}$. (e) Exciton-polariton band structure from simulation and theory with phonon coupling $2\gamma_0$ with similar parameters used in (b). We used  $\gamma_{0}= 5.85 \times 10^{-4} $ and $1.46 \times 10^{-4}$ a.u.  for figure a and b respectively. Further we use $\Omega = 3900$  cm$^{-1}$, $N = 40001 $ for figure a and $N = 30001$ for figure b, $\tau = 0$, $\omega_0 = 2.58$ eV, and $\varepsilon_0 = 3.2$ eV.}

\label{fig2}
\end{figure}
 
With the success in obtaining the phonon-modified polariton bands (which we refer to as polaron polariton bands) using our analytical model, in Fig.~\ref{fig2} we use it to obtain the group velocities and provide insights into the coherent propagation of polaron-polaritons. In the absence of phonons, a coherent superposition of neighboring wave vectors in the momentum space, e.g.  $|\Psi \rangle = \lim_{F \delta k \rightarrow 0}\frac{1}{\sqrt{F}}\sum_{n = 1}^{F}|k + n \delta k\rangle$ in an extended system, leads to coherent ballistic propagation with a group velocity equal to the slope of the band structure $dE/dk$, known as the group velocity. On the other hand, in the presence of phonons, phonon-induced decoherence leads to incoherent diffusive motion. Therefore, it is expected that exciton-polaritons, depending on the extent of their material character, will lead to short-time (for times  less than the decoherence lifetime) coherent ballistic motion with group velocities matching the exciton-polariton dispersion. Experimental results, however, indicate that  exciton-polaritons with significantly high excitonic character (up to $\sim$50\% excitonic) show long-lived coherent ballistic motion (up to hundreds of femtoseconds) with group velocities lower than the slopes of the exciton-polariton band structure. Our theory provides a plausible explanation for this twofold mystery and provides new microscopic insights into this extraordinary phenomenon. 

Fig.~\ref{fig2}a-b presents the polariton group velocity obtained from our analytical model (solid lines), comparing it to the group velocities obtained by performing direct quantum dynamical simulations (filled circles) at two different phonon frequencies and various phonon couplings. Fig.~\ref{fig2}c-d presents time-dependent excitonic density $\rho_n (t) = \langle \langle \Psi (t) | \hat{X}_n^\dagger  \hat{X}_n|  \Psi (t)\rangle  \rangle_\mathrm{MFE}$ in the presence and in the absence of phonon couplings.  To perform these simulations, we have prepared the initial exciton-polariton wavefunction as a linear combination of polariton states within an energy window $\Delta E$ centered at an excitation energy $E_0$,  such that $|\Psi (0)\rangle = \sum c_j |E_j\rangle$ with $E_0 -\Delta E/2  < E_j < E_0 + \Delta E/2$ and  $|E_j\rangle$ as the eigenstates of $\hat{H}_{EP}$. In both cases, we observe a ballistic propagation suggested by the linear expansion of the wavefront in time, with the latter propagating relatively slowly as the group velocities presented in Fig.~\ref{fig2}a-b. We extract the group velocities from these wavefronts, which are presented in Fig.~\ref{fig2}a-b (filled circles) and are compared to the predictions of our analytical model. 

Overall, the results presented here clearly illustrate the applicability of our analytical model and quasi-band structures introduced here for understanding the polariton propagation. At higher phonon frequencies, the vibronic structure in the dispersion directly results in an oscillatory behavior in the group velocity with troughs separated by the phonon frequency $\omega$. At lower phonon frequencies, such as in Fig.~\ref{fig2}b, the oscillatory structure is almost absent as the peaks in the analytical theory pack closer.  Fig.~\ref{fig2}e presents the angle-resolved spectra at $\omega$ = 360 cm$^{-1}$  where the vibronic peaks are no longer visible due to the finite linewidth of the optical spectra. Therefore, even though the vibronic structure is not visible in polaritonic spectra, it results in a renormalization of the group velocity.  This phenomenon has been observed experimentally~\cite{XuNC2023, BalasubrahmaniyamNM2023}, with our theory providing a clear theoretical explanation. 

In both scenarios, however, the observed group velocities are always lower, due to the formation of the polaron-polariton (quasi) bands that have flatter slopes, due to the contribution of the flat effective exciton bands $\hat{Y}_{k,m}$, compared to the bare exciton-polariton dispersion. This renormalization of the exciton-polariton group velocity is induced by the presence of phonons in materials, and even at low phonon frequencies, where the vibronic structure in the angle-resolved spectra may be hidden due to various sources of dissipation (such as cavity loss), the quasi-bands lead to the renormalization of the group velocity.
%That is, phonons induce a renormalization of the exciton-polariton group velocity. Note that at  low phonon frequencies, while the vibronic structure may get hidden in the angle-resolved spectra, it still dictates the propagation of the exciton-polariton.  
Overall, our theoretical model correctly captures the complex ballistic propagation of exciton-polaritons in the presence of phonon interactions and introduces a quasi-band picture that can be adopted to describe and understand the coherent propagation of polaron-polaritons.   

%In pure excitonic systems, the effect of phonons is threefold: phonons induce decoherence, turning coherent ballistic propagation into diffusive incoherent motion, effectively suppressing the hopping term $\tau$, making the short-time coherent motion much slower, and breaking the translational symmetry.

Importantly, our work also suggests a microscopic explanation for the relatively long-lived coherent propagation of exciton-polaritons with high exciton character~\cite{XuNC2023, BalasubrahmaniyamNM2023, PandyaAdvS2022, SandikNatMat2024}. We hypothesize that the origin of this extraordinary effect is the block diagonal nature of Eq.~\ref{effective} where photon modes $\hat{a}_k^{\dagger}$ couple to a particular set of {\it effective} reciprocal (phonon-dressed) excitons $\hat{Y}_{k,m}$ with matching $k$, defined in Eq.~\ref{effective}. To clearly understand the ramifications of this, consider first a bare excitonic system coupled with phonons under laser driving $\mathcal{E}(t)$ that target a subspace $\mathcal{K}$ in reciprocal space can be written as 
\begin{align}
\hat{H}_\mathrm{X} +  \hat{H}_\mathrm{laser}  &= \sum_k \hat{X}_{k}^{\dagger}  \hat{X}_k \epsilon_k +  \frac{\gamma}{\sqrt{2\omega}}\sum_{k,q} \hat{X}_{k+q}^{\dagger}  \hat{X}_k (\hat{b}_{q} + \hat{b}_{-q}^\dagger)  \nonumber \\
&+ \sum_k \hat{b}_{k}^{\dagger}\hat{b}_{k}\omega + \mathcal{E}(t) \sum_{k \in \mathcal{K}} (\hat{X}_{k}^{\dagger} + \hat{X}_{k}) 
\end{align}
where $\epsilon_k = \epsilon_0 + 2\tau \cos(k\cdot a)$ for choice of nearest neighbor interactions made here. Importantly, despite a laser exclusively targeting the subspace  $\mathcal{K}$, the population leaks out to the subspace $\mathcal{M} = \mathds{1} - \mathcal{K}$ via the phonon-induced scattering term $\frac{\gamma}{\sqrt{2\omega}}\sum_{k,q} \hat{X}_{k+q}^{\dagger}  \hat{X}_k (\hat{b}_{q} + \hat{b}_{-q}^\dagger)$. In contrast, inside an optical cavity, following our analytical model in Eq.~\ref{HFeff}, a driven light-matter hybrid system can be modeled as~\cite{HerreraPRA2017}
\begin{align}
  \hat{H}_\mathrm{LM} + \hat{H}_\mathrm{laser}  &= \hat{H}'_\mathrm{F} + \mathcal{E}(t) \sum_{k \in \mathcal{K}} (\hat{a}_{k}^{\dagger} + \hat{a}_{k}) \nonumber \\
  &\approx \sum_{k \in \mathcal{K}} \Big[\mathcal{\hat{H}}_k + \mathcal{E}(t) (\hat{a}_{k}^{\dagger} + \hat{a}_{k}) \Big] +  \sum_{k \in \mathcal{M}} \mathcal{\hat{H}}_k,
\end{align}
such that the subspace $\mathcal{M}$ and $\mathcal{K}$ now remain decoupled. Therefore, light-matter interaction also plays a crucial role in suppressing phonon-induced scattering in the reciprocal space, allowing for relatively long-lived ballistic motion in the time scale of hundreds of femtoseconds. 

%Meanwhile, new polaron-polariton (quasi) bands borrow the curvature of the original photonic band and assume a large group velocity in the presence of phonon interactions.  As a result, coherent propagation can be observed in the presence of phonons (at short times) but with group velocities that match the slopes of the polaron-polariton bands instead of the exciton polariton ones.  

In summary, we developed a convenient theoretical framework to understand and predict the angle-resolved polariton spectra in the presence of phonon interactions. Starting from a microscopic Hamiltonian describing the interactions between phonons, excitons, and photons inside an optical cavity, we develop an analytical model that accurately predicts the complex angle-resolved polariton spectra and the group velocities of coherently propagating exciton-polaritons. We derive this analytical model by describing the phonons as time-periodic fields that are non-perturbatively interacting with exciton-polaritons and quantize them using the Floquet formalism that is typically used to describe laser-matter interactions. Note that despite the classical treatment of phonons within the mixed quantum-classical framework, the vibronic structure obtained is expected to be reasonably accurate given its success in  various model systems~\cite{EgorovJPCB1999, PetitJCP2014, Mondal2023JCP, ProvazzaJCTC2018}. We show that our theory can accurately predict the group velocity of the coherent ballistic propagation of exciton-polaritons in the presence of phonon interactions, i.e., polaron-polaritons, and provides a new perspective into the observed renormalization of polaritonic group velocity and their long-lived coherent nature in recent experiments.

{\bf Acknowledgments}. This work was supported by the Texas A\&M startup
funds. This work used TAMU FASTER at the Texas A\&M University through allocation  PHY230021 from the Advanced Cyberinfrastructure Coordination Ecosystem: Services \& Support (ACCESS) program, which is supported by National Science Foundation grants \#2138259, \#2138286, \#2138307, \#2137603, and \#2138296. A.M. appreciates inspiring discussions with Micheal Taylor, Daniel Tabor, Milan Delor, Pengfei Huo, David R. Reichman, Wenxiang Ying, Dipti Jasrasaria and Haimi Nguyen. The authors appreciate discussions with Pritha Ghosh and Sachith Wickramasinghe.

}
\bibliography{bib.bib}

\begin{thebibliography}{10}
\expandafter\ifx\csname url\endcsname\relax
  \def\url#1{\texttt{#1}}\fi
\expandafter\ifx\csname urlprefix\endcsname\relax\def\urlprefix{URL }\fi
\providecommand{\bibinfo}[2]{#2}
\providecommand{\eprint}[2][]{\url{#2}}

\bibitem{LiARPC2022}
\bibinfo{author}{Li, T.~E.}, \bibinfo{author}{Cui, B.}, \bibinfo{author}{Subotnik, J.~E.} \& \bibinfo{author}{Nitzan, A.}
\newblock \bibinfo{title}{Molecular polaritonics: Chemical dynamics under strong light--matter coupling}.
\newblock \emph{\bibinfo{journal}{Annual review of physical chemistry}} \textbf{\bibinfo{volume}{73}}, \bibinfo{pages}{43--71} (\bibinfo{year}{2022}).

\bibitem{SnokePhyToday2010}
\bibinfo{author}{Snoke, D.} \& \bibinfo{author}{Littlewood, P.}
\newblock \bibinfo{title}{Polariton condensates}.
\newblock \emph{\bibinfo{journal}{Physics Today}} \textbf{\bibinfo{volume}{63}}, \bibinfo{pages}{42--47} (\bibinfo{year}{2010}).

\bibitem{DengRMP2010}
\bibinfo{author}{Deng, H.}, \bibinfo{author}{Haug, H.} \& \bibinfo{author}{Yamamoto, Y.}
\newblock \bibinfo{title}{Exciton-polariton bose-einstein condensation}.
\newblock \emph{\bibinfo{journal}{Rev. Mod. Phys.}} \textbf{\bibinfo{volume}{82}}, \bibinfo{pages}{1489--1537} (\bibinfo{year}{2010}).

\bibitem{LidzeyPRL1999}
\bibinfo{author}{Lidzey, D.~G.} \emph{et~al.}
\newblock \bibinfo{title}{Room temperature polariton emission from strongly coupled organic semiconductor microcavities}.
\newblock \emph{\bibinfo{journal}{Phys. Rev. Lett.}} \textbf{\bibinfo{volume}{82}}, \bibinfo{pages}{3316--3319} (\bibinfo{year}{1999}).

\bibitem{SandikNatMat2024}
\bibinfo{author}{Sandik, G.}, \bibinfo{author}{Feist, J.}, \bibinfo{author}{García-Vidal, F.~J.} \& \bibinfo{author}{Schwartz, T.}
\newblock \bibinfo{title}{Cavity-enhanced energy transport in molecular systems}.
\newblock \emph{\bibinfo{journal}{Nature Materials}}  (\bibinfo{year}{2024}).

\bibitem{KowalewskiJCP2016}
\bibinfo{author}{Kowalewski, M.}, \bibinfo{author}{Bennett, K.} \& \bibinfo{author}{Mukamel, S.}
\newblock \bibinfo{title}{Non-adiabatic dynamics of molecules in optical cavities}.
\newblock \emph{\bibinfo{journal}{The Journal of Chemical Physics}} \textbf{\bibinfo{volume}{144}}, \bibinfo{pages}{054309} (\bibinfo{year}{2016}).

\bibitem{MandalCR2022}
\bibinfo{author}{Mandal, A.} \emph{et~al.}
\newblock \bibinfo{title}{Theoretical advances in polariton chemistry and molecular cavity quantum electrodynamics}.
\newblock \emph{\bibinfo{journal}{chemrxiv-2022-g9lr7}}  (\bibinfo{year}{2022}).

\bibitem{KeelingARPC2020}
\bibinfo{author}{Keeling, J.} \& \bibinfo{author}{Kéna-Cohen, S.}
\newblock \bibinfo{title}{Bose–einstein condensation of exciton-polaritons in organic microcavities}.
\newblock \emph{\bibinfo{journal}{Annual Review of Physical Chemistry}} \textbf{\bibinfo{volume}{71}}, \bibinfo{pages}{435--459} (\bibinfo{year}{2020}).

\bibitem{YueningACSP2024}
\bibinfo{author}{Fan, Y.} \emph{et~al.}
\newblock \bibinfo{title}{High efficiency of exciton-polariton lasing in a 2d multilayer structure}.
\newblock \emph{\bibinfo{journal}{ACS photonics}} \textbf{\bibinfo{volume}{11}}, \bibinfo{pages}{2722--2728} (\bibinfo{year}{2024}).

\bibitem{KockumNRP2019}
\bibinfo{author}{Frisk~Kockum, A.}, \bibinfo{author}{Miranowicz, A.}, \bibinfo{author}{De~Liberato, S.}, \bibinfo{author}{Savasta, S.} \& \bibinfo{author}{Nori, F.}
\newblock \bibinfo{title}{Ultrastrong coupling between light and matter}.
\newblock \emph{\bibinfo{journal}{Nature Reviews Physics}} \textbf{\bibinfo{volume}{1}}, \bibinfo{pages}{19--40} (\bibinfo{year}{2019}).

\bibitem{XuNN2025}
\bibinfo{author}{Xu, D.} \emph{et~al.}
\newblock \bibinfo{title}{Spatiotemporal imaging of nonlinear optics in van der waals waveguides}.
\newblock \emph{\bibinfo{journal}{Nature Nanotechnology}} \bibinfo{pages}{1--7} (\bibinfo{year}{2025}).

\bibitem{Wurdack2023NC}
\bibinfo{author}{Wurdack, M.} \emph{et~al.}
\newblock \bibinfo{title}{Negative-mass exciton polaritons induced by dissipative light-matter coupling in an atomically thin semiconductor}.
\newblock \emph{\bibinfo{journal}{Nature Communications}} \textbf{\bibinfo{volume}{14}}, \bibinfo{pages}{1026} (\bibinfo{year}{2023}).

\bibitem{Matthias2021NC}
\bibinfo{author}{Wurdack, M.} \emph{et~al.}
\newblock \bibinfo{title}{Motional narrowing, ballistic transport, and trapping of room-temperature exciton polaritons in an atomically-thin semiconductor}.
\newblock \emph{\bibinfo{journal}{Nature communications}} \textbf{\bibinfo{volume}{12}}, \bibinfo{pages}{5366} (\bibinfo{year}{2021}).

\bibitem{SuyabatmazJCP2023}
\bibinfo{author}{Suyabatmaz, E.} \& \bibinfo{author}{Ribeiro, R.~F.}
\newblock \bibinfo{title}{Vibrational polariton transport in disordered media}.
\newblock \emph{\bibinfo{journal}{The Journal of Chemical Physics}} \textbf{\bibinfo{volume}{159}}, \bibinfo{pages}{034701} (\bibinfo{year}{2023}).

\bibitem{LiuNP2015}
\bibinfo{author}{Liu, X.} \emph{et~al.}
\newblock \bibinfo{title}{Strong light--matter coupling in two-dimensional atomic crystals}.
\newblock \emph{\bibinfo{journal}{Nature Photonics}} \textbf{\bibinfo{volume}{9}}, \bibinfo{pages}{30--34} (\bibinfo{year}{2015}).

\bibitem{kenaNP2010}
\bibinfo{author}{K{\'e}na-Cohen, S.} \& \bibinfo{author}{Forrest, S.}
\newblock \bibinfo{title}{Room-temperature polariton lasing in an organic single-crystal microcavity}.
\newblock \emph{\bibinfo{journal}{Nature Photonics}} \textbf{\bibinfo{volume}{4}}, \bibinfo{pages}{371--375} (\bibinfo{year}{2010}).

\bibitem{StegerPRB2013}
\bibinfo{author}{Steger, M.} \emph{et~al.}
\newblock \bibinfo{title}{Long-range ballistic motion and coherent flow of long-lifetime polaritons}.
\newblock \emph{\bibinfo{journal}{Phys. Rev. B}} \textbf{\bibinfo{volume}{88}}, \bibinfo{pages}{235314} (\bibinfo{year}{2013}).

\bibitem{XuNC2023}
\bibinfo{author}{Xu, D.} \emph{et~al.}
\newblock \bibinfo{title}{Ultrafast imaging of polariton propagation and interactions}.
\newblock \emph{\bibinfo{journal}{Nature Communications}} \textbf{\bibinfo{volume}{14}}, \bibinfo{pages}{4082--4089} (\bibinfo{year}{2023}).

\bibitem{BalasubrahmaniyamNM2023}
\bibinfo{author}{Balasubrahmaniyam, M.} \emph{et~al.}
\newblock \bibinfo{title}{From enhanced diffusion to ultrafast ballistic motion of hybrid light--matter excitations}.
\newblock \emph{\bibinfo{journal}{Nature Materials}} \textbf{\bibinfo{volume}{22}}, \bibinfo{pages}{338--344} (\bibinfo{year}{2023}).

\bibitem{sanvittoNM2016}
\bibinfo{author}{Sanvitto, D.} \& \bibinfo{author}{K{\'e}na-Cohen, S.}
\newblock \bibinfo{title}{The road towards polaritonic devices}.
\newblock \emph{\bibinfo{journal}{Nature materials}} \textbf{\bibinfo{volume}{15}}, \bibinfo{pages}{1061--1073} (\bibinfo{year}{2016}).

\bibitem{FerreiraPRM2022}
\bibinfo{author}{Ferreira, B.}, \bibinfo{author}{Rosati, R.} \& \bibinfo{author}{Malic, E.}
\newblock \bibinfo{title}{Microscopic modeling of exciton-polariton diffusion coefficients in atomically thin semiconductors}.
\newblock \emph{\bibinfo{journal}{Physical Review Materials}} \textbf{\bibinfo{volume}{6}}, \bibinfo{pages}{034008} (\bibinfo{year}{2022}).

\bibitem{DhamijaCPR2024}
\bibinfo{author}{Dhamija, S.} \& \bibinfo{author}{Son, M.}
\newblock \bibinfo{title}{Mapping the dynamics of energy relaxation in exciton--polaritons using ultrafast two-dimensional electronic spectroscopy}.
\newblock \emph{\bibinfo{journal}{Chemical Physics Reviews}} \textbf{\bibinfo{volume}{5}} (\bibinfo{year}{2024}).

\bibitem{KadeCR2020}
\bibinfo{author}{Head-Marsden, K.}, \bibinfo{author}{Flick, J.}, \bibinfo{author}{Ciccarino, C.~J.} \& \bibinfo{author}{Narang, P.}
\newblock \bibinfo{title}{Quantum information and algorithms for correlated quantum matter}.
\newblock \emph{\bibinfo{journal}{Chemical Reviews}} \textbf{\bibinfo{volume}{121}}, \bibinfo{pages}{3061--3120} (\bibinfo{year}{2020}).

\bibitem{Reineker1984Book}
\bibinfo{author}{Reineker, P.}
\newblock \emph{\bibinfo{title}{Exciton dynamics in molecular crystals and aggregates}}, vol.~\bibinfo{volume}{94} (\bibinfo{publisher}{Springer}, \bibinfo{year}{1982}).

\bibitem{TroisiPRL2006}
\bibinfo{author}{Troisi, A.} \& \bibinfo{author}{Orlandi, G.}
\newblock \bibinfo{title}{Charge-transport regime of crystalline organic semiconductors: Diffusion limited by thermal off-diagonal electronic disorder}.
\newblock \emph{\bibinfo{journal}{Physical review letters}} \textbf{\bibinfo{volume}{96}}, \bibinfo{pages}{086601} (\bibinfo{year}{2006}).

\bibitem{Jonathan2020PRX}
\bibinfo{author}{Fetherolf, J.~H.}, \bibinfo{author}{Gole{\v{z}}, D.} \& \bibinfo{author}{Berkelbach, T.~C.}
\newblock \bibinfo{title}{A unification of the holstein polaron and dynamic disorder pictures of charge transport in organic crystals}.
\newblock \emph{\bibinfo{journal}{Physical Review X}} \textbf{\bibinfo{volume}{10}}, \bibinfo{pages}{021062} (\bibinfo{year}{2020}).

\bibitem{Linjun2011JCP}
\bibinfo{author}{Wang, L.}, \bibinfo{author}{Beljonne, D.}, \bibinfo{author}{Chen, L.} \& \bibinfo{author}{Shi, Q.}
\newblock \bibinfo{title}{Mixed quantum-classical simulations of charge transport in organic materials: Numerical benchmark of the su-schrieffer-heeger model}.
\newblock \emph{\bibinfo{journal}{The Journal of chemical physics}} \textbf{\bibinfo{volume}{134}} (\bibinfo{year}{2011}).

\bibitem{YarkonyJCP1977}
\bibinfo{author}{Yarkony, D.~R.} \& \bibinfo{author}{Silbey, R.}
\newblock \bibinfo{title}{Variational approach to exciton transport in molecular crystals}.
\newblock \emph{\bibinfo{journal}{The Journal of Chemical Physics}} \textbf{\bibinfo{volume}{67}}, \bibinfo{pages}{5818--5827} (\bibinfo{year}{1977}).

\bibitem{BalasubrahmaniyamPRB2021}
\bibinfo{author}{Balasubrahmaniyam, M.}, \bibinfo{author}{Genet, C.} \& \bibinfo{author}{Schwartz, T.}
\newblock \bibinfo{title}{Coupling and decoupling of polaritonic states in multimode cavities}.
\newblock \emph{\bibinfo{journal}{Phys. Rev. B}} \textbf{\bibinfo{volume}{103}}, \bibinfo{pages}{L241407} (\bibinfo{year}{2021}).

\bibitem{Krupp2024}
\bibinfo{author}{Krupp, N.}, \bibinfo{author}{Groenhof, G.} \& \bibinfo{author}{Oriol, V.}
\newblock \bibinfo{title}{Quantum dynamics simulation of exciton-polariton transport}.
\newblock \emph{\bibinfo{journal}{arXiv:physics.chem-ph/2411.23739}}  (\bibinfo{year}{2024}).

\bibitem{BerghuisACSp2022}
\bibinfo{author}{Berghuis, A.~M.} \emph{et~al.}
\newblock \bibinfo{title}{Controlling exciton propagation in organic crystals through strong coupling to plasmonic nanoparticle arrays}.
\newblock \emph{\bibinfo{journal}{ACS Photonics}} \textbf{\bibinfo{volume}{9}}, \bibinfo{pages}{2263--2272} (\bibinfo{year}{2022}).

\bibitem{PandyaAdvS2022}
\bibinfo{author}{Pandya, R.} \emph{et~al.}
\newblock \bibinfo{title}{Tuning the coherent propagation of organic exciton-polaritons through dark state delocalization}.
\newblock \emph{\bibinfo{journal}{Advances Science}} \textbf{\bibinfo{volume}{9}}, \bibinfo{pages}{2105569} (\bibinfo{year}{2022}).

\bibitem{LiuACSP2020}
\bibinfo{author}{Liu, B.}, \bibinfo{author}{Menon, V.~M.} \& \bibinfo{author}{Sfeir, M.~Y.}
\newblock \bibinfo{title}{The role of long-lived excitons in the dynamics of strongly coupled molecular polaritons}.
\newblock \emph{\bibinfo{journal}{ACS Photonics}} \textbf{\bibinfo{volume}{7}}, \bibinfo{pages}{2292--2301} (\bibinfo{year}{2020}).

\bibitem{SokolovskiiNatCom2023}
\bibinfo{author}{Sokolovskii, I.}, \bibinfo{author}{Tichauer, R.~H.}, \bibinfo{author}{Morozov, D.}, \bibinfo{author}{Feist, J.} \& \bibinfo{author}{Groenhof, G.}
\newblock \bibinfo{title}{Multi-scale molecular dynamics simulations of enhanced energy transfer in organic molecules under strong coupling}.
\newblock \emph{\bibinfo{journal}{Nature Communications}} \textbf{\bibinfo{volume}{14}} (\bibinfo{year}{2023}).

\bibitem{TichauerAS2023}
\bibinfo{author}{Tichauer, R.~H.}, \bibinfo{author}{Sokolovskii, I.} \& \bibinfo{author}{Groenhof, G.}
\newblock \bibinfo{title}{Tuning the coherent propagation of organic exciton-polaritons through the cavity q-factor}.
\newblock \emph{\bibinfo{journal}{Advanced Science}} \textbf{\bibinfo{volume}{10}}, \bibinfo{pages}{2302650} (\bibinfo{year}{2023}).

\bibitem{ChngNL2025}
\bibinfo{author}{Chng, B. X.~K.}, \bibinfo{author}{Mondal, M.~E.}, \bibinfo{author}{Ying, W.} \& \bibinfo{author}{Huo, P.}
\newblock \bibinfo{title}{Quantum dynamics simulations of exciton polariton transport}.
\newblock \emph{\bibinfo{journal}{Nano Letters}} \textbf{\bibinfo{volume}{0}}, \bibinfo{pages}{null} (\bibinfo{year}{2025}).

\bibitem{Ying2024}
\bibinfo{author}{Ying, W.}, \bibinfo{author}{Chng, B. X.~K.} \& \bibinfo{author}{Huo, P.}
\newblock \bibinfo{title}{Microscopic theory of polariton group velocity renormalization}.
\newblock \emph{\bibinfo{journal}{arXiv:quant-ph/2411.08288}}  (\bibinfo{year}{2024}).

\bibitem{TutunnikovArxiv2024}
\bibinfo{author}{Tutunnikov, I.}, \bibinfo{author}{Qutubuddin, M.}, \bibinfo{author}{Sadeghpour, H.} \& \bibinfo{author}{Cao, J.}
\newblock \bibinfo{title}{Characterization of polariton dynamics in a multimode cavity: Noise-enhanced ballistic expansion}.
\newblock \emph{\bibinfo{journal}{arXiv preprint arXiv:2410.11051}}  (\bibinfo{year}{2024}).

\bibitem{AroeiraNp2024}
\bibinfo{author}{Aroeira, G.~J.}, \bibinfo{author}{Kairys, K.~T.} \& \bibinfo{author}{Ribeiro, R.~F.}
\newblock \bibinfo{title}{Coherent transient exciton transport in disordered polaritonic wires}.
\newblock \emph{\bibinfo{journal}{Nanophotonics}} \textbf{\bibinfo{volume}{13}}, \bibinfo{pages}{2553--2564} (\bibinfo{year}{2024}).

\bibitem{EngelhardtPRL2023}
\bibinfo{author}{Engelhardt, G.} \& \bibinfo{author}{Cao, J.}
\newblock \bibinfo{title}{Polariton localization and dispersion properties of disordered quantum emitters in multimode microcavities}.
\newblock \emph{\bibinfo{journal}{Phys. Rev. Lett.}} \textbf{\bibinfo{volume}{130}}, \bibinfo{pages}{213602} (\bibinfo{year}{2023}).

\bibitem{MandalNL2023}
\bibinfo{author}{Mandal, A.} \emph{et~al.}
\newblock \bibinfo{title}{Microscopic theory of multimode polariton dispersion in multilayered materials}.
\newblock \emph{\bibinfo{journal}{Nano Letters}} \textbf{\bibinfo{volume}{23}}, \bibinfo{pages}{4082--4089} (\bibinfo{year}{2023}).

\bibitem{Hoffmann2020JCP}
\bibinfo{author}{Hoffmann, N.~M.}, \bibinfo{author}{Lacombe, L.}, \bibinfo{author}{Rubio, A.} \& \bibinfo{author}{Maitra, N.~T.}
\newblock \bibinfo{title}{Effect of many modes on self-polarization and photochemical suppression in cavities}.
\newblock \emph{\bibinfo{journal}{The Journal of Chemical Physics}} \textbf{\bibinfo{volume}{153}} (\bibinfo{year}{2020}).

\bibitem{LiPRA2018}
\bibinfo{author}{Li, T.~E.} \emph{et~al.}
\newblock \bibinfo{title}{Mixed quantum-classical electrodynamics: Understanding spontaneous decay and zero-point energy}.
\newblock \emph{\bibinfo{journal}{Physical Review A}} \textbf{\bibinfo{volume}{97}}, \bibinfo{pages}{032105} (\bibinfo{year}{2018}).

\bibitem{EgorovJPCB1999}
\bibinfo{author}{Egorov, S.~A.}, \bibinfo{author}{Rabani, E.} \& \bibinfo{author}{Berne, B.~J.}
\newblock \bibinfo{title}{On the adequacy of mixed quantum-classical dynamics in condensed phase systems}.
\newblock \emph{\bibinfo{journal}{The Journal of Physical Chemistry B}} \textbf{\bibinfo{volume}{103}}, \bibinfo{pages}{10978--10991} (\bibinfo{year}{1999}).

\bibitem{PetitJCP2014}
\bibinfo{author}{Petit, A.~S.} \& \bibinfo{author}{Subotnik, J.~E.}
\newblock \bibinfo{title}{How to calculate linear absorption spectra with lifetime broadening using fewest switches surface hopping trajectories: A simple generalization of ground-state kubo theory}.
\newblock \emph{\bibinfo{journal}{The Journal of chemical physics}} \textbf{\bibinfo{volume}{141}} (\bibinfo{year}{2014}).

\bibitem{DanielCS2020}
\bibinfo{author}{Polak, D.} \emph{et~al.}
\newblock \bibinfo{title}{Manipulating molecules with strong coupling: harvesting triplet excitons in organic exciton microcavities}.
\newblock \emph{\bibinfo{journal}{Chemical science}} \textbf{\bibinfo{volume}{11}}, \bibinfo{pages}{343--354} (\bibinfo{year}{2020}).

\bibitem{DonghaiPRL2022}
\bibinfo{author}{Li, D.} \emph{et~al.}
\newblock \bibinfo{title}{Hybridized exciton-photon-phonon states in a transition metal dichalcogenide van der waals heterostructure microcavity}.
\newblock \emph{\bibinfo{journal}{Physical Review Letters}} \textbf{\bibinfo{volume}{128}}, \bibinfo{pages}{087401} (\bibinfo{year}{2022}).

\bibitem{AkimovJCTC2013}
\bibinfo{author}{Akimov, A.~V.} \& \bibinfo{author}{Prezhdo, O.~V.}
\newblock \bibinfo{title}{The pyxaid program for non-adiabatic molecular dynamics in condensed matter systems}.
\newblock \emph{\bibinfo{journal}{Journal of chemical theory and computation}} \textbf{\bibinfo{volume}{9}}, \bibinfo{pages}{4959--4972} (\bibinfo{year}{2013}).

\bibitem{YamijalaJPCA2021}
\bibinfo{author}{Yamijala, S.~S.} \& \bibinfo{author}{Huo, P.}
\newblock \bibinfo{title}{Direct nonadiabatic simulations of the photoinduced charge transfer dynamics}.
\newblock \emph{\bibinfo{journal}{The Journal of Physical Chemistry A}} \textbf{\bibinfo{volume}{125}}, \bibinfo{pages}{628--635} (\bibinfo{year}{2021}).

\bibitem{Shirley1965PR}
\bibinfo{author}{Shirley, J.~H.}
\newblock \bibinfo{title}{Solution of the schr{\"o}dinger equation with a hamiltonian periodic in time}.
\newblock \emph{\bibinfo{journal}{Physical Review}} \textbf{\bibinfo{volume}{138}}, \bibinfo{pages}{B979} (\bibinfo{year}{1965}).

\bibitem{TaylorCPR2024}
\bibinfo{author}{Taylor, M.}, \bibinfo{author}{Mandal, A.} \& \bibinfo{author}{Huo, P.}
\newblock \bibinfo{title}{Light-matter interaction hamiltonians in cavity quantum electrodynamics}.
\newblock \emph{\bibinfo{journal}{Chemical Physics Reviews}}  (\bibinfo{year}{2025}).

\bibitem{spano2015optical}
\bibinfo{author}{Spano, F.}
\newblock \bibinfo{title}{Optical microcavities enhance the exciton coherence length and eliminate vibronic coupling in j-aggregates}.
\newblock \emph{\bibinfo{journal}{The Journal of Chemical Physics}} \textbf{\bibinfo{volume}{142}} (\bibinfo{year}{2015}).

\bibitem{MazzaPRB2013}
\bibinfo{author}{Mazza, L.}, \bibinfo{author}{K{\'e}na-Cohen, S.}, \bibinfo{author}{Michetti, P.} \& \bibinfo{author}{La~Rocca, G.~C.}
\newblock \bibinfo{title}{Microscopic theory of polariton lasing via vibronically assisted scattering}.
\newblock \emph{\bibinfo{journal}{Physical Review B—Condensed Matter and Materials Physics}} \textbf{\bibinfo{volume}{88}}, \bibinfo{pages}{075321} (\bibinfo{year}{2013}).

\bibitem{FontanesiPRB2009}
\bibinfo{author}{Fontanesi, L.}, \bibinfo{author}{Mazza, L.} \& \bibinfo{author}{La~Rocca, G.~C.}
\newblock \bibinfo{title}{Organic-based microcavities with vibronic progressions: Linear spectroscopy}.
\newblock \emph{\bibinfo{journal}{Physical Review B—Condensed Matter and Materials Physics}} \textbf{\bibinfo{volume}{80}}, \bibinfo{pages}{235313} (\bibinfo{year}{2009}).

\bibitem{HerreraPRA2017}
\bibinfo{author}{Herrera, F.} \& \bibinfo{author}{Spano, F.~C.}
\newblock \bibinfo{title}{Absorption and photoluminescence in organic cavity qed}.
\newblock \emph{\bibinfo{journal}{Physical Review A}} \textbf{\bibinfo{volume}{95}}, \bibinfo{pages}{053867} (\bibinfo{year}{2017}).

\bibitem{Mondal2023JCP}
\bibinfo{author}{Mondal, M.~E.} \emph{et~al.}
\newblock \bibinfo{title}{Quantum dynamics simulations of the 2d spectroscopy for exciton polaritons}.
\newblock \emph{\bibinfo{journal}{The Journal of Chemical Physics}} \textbf{\bibinfo{volume}{159}} (\bibinfo{year}{2023}).

\bibitem{ProvazzaJCTC2018}
\bibinfo{author}{Provazza, J.}, \bibinfo{author}{Segatta, F.}, \bibinfo{author}{Garavelli, M.} \& \bibinfo{author}{Coker, D.~F.}
\newblock \bibinfo{title}{Semiclassical path integral calculation of nonlinear optical spectroscopy}.
\newblock \emph{\bibinfo{journal}{Journal of chemical theory and computation}} \textbf{\bibinfo{volume}{14}}, \bibinfo{pages}{856--866} (\bibinfo{year}{2018}).

\end{thebibliography}
\bibliographystyle{naturemag}

\end{document}